\documentclass[twocolumn,aps,showpacs,pre,amsmath]{revtex4}

\usepackage{amssymb}
\usepackage{graphicx}

\begin{document}

\title{Conduction gap in graphene strain junctions: direction dependence}

\author{M. Chung Nguyen$^1$, V. Hung Nguyen$^{1,2,3}$\footnote{E-mail: hung@iop.vast.ac.vn}, Huy-Viet Nguyen$^3$ and P. Dollfus$^1$} \address{$^1$Institut d'Electronique Fondamentale, UMR8622, CNRS, Universit$\acute{e}$ Paris Sud, 91405 Orsay, France \\ $^2$L$_-$Sim, SP2M, UMR-E CEA/UJF-Grenoble 1, INAC, 38054 Grenoble, France \\ $^3$Center for Computational Physics, Institute of Physics, Vietnam Academy of Science and Technology, P.O. Box 429 Bo Ho, 10000 Hanoi, Vietnam}

\begin{abstract}
 It has been shown in a recent study [Nguyen et al., Nanotechnol. \textbf{25}, 165201 (2014)] that unstrained/strained graphene junctions are promising candidates to improve the performance of graphene transistors that is usually hindered by the gapless nature of graphene. Although the energy bandgap of strained graphene still remains zero, the shift of Dirac points in the \textbf{\emph{k}}-space due to strain-induced deformation of graphene lattice can lead to the appearance of a finite conduction gap of several hundreds meV in strained junctions with a strain of only a few percent. However, since it depends essentially on the magnitude of Dirac point shift, this conduction gap strongly depends on the direction of applied strain and the transport direction. In this work, a systematic study of conduction gap properties with respect to these quantities is presented and the results are carefully analyzed. Our study provides useful information for further investigations to exploit graphene strained junctions in electronic applications.
\end{abstract}

\pacs{xx.xx.xx, yy.yy.yy, zz.zz.zz}
\maketitle

\section{Introduction}

In spite of being an attractive material with excellent electronic properties \cite{ahcn09},  practical applications of graphene as in conventional semiconductor devices are still questionable due to its gapless nature. In particular, the ON/OFF current ratio is low while the saturation of current is poor in pristine graphene transistors \cite{schw10}. Many efforts of bandgap engineering in graphene \cite{yhan07,khar11,lher13,jbai10,zhan09} have been made to solve these issues. The pioneer technique proposed \cite{yhan07} is to cut 2D graphene sheets into 1D narrow nanoribons. In 2D graphene sheets, some options as Bernal-stacking of graphene on hexagonal boron nitride substrate \cite{khar11}, nitrogen-doped graphene \cite{lher13}, graphene nanomesh lattice \cite{jbai10,berr13} and Bernal-stacking bilayer graphene \cite{zhan09} have been explored. However, the possibility to open a sizable bandgap in graphene as large as those of standard semiconductors is still very unlikely. In particular, it requires a very good control of lattice geometry and edge disorder in narrow graphene nanoribbons (GNRs) \cite{quer08} and in graphene nanomesh lattices \cite{hung13}, while the bandgap opening in bilayer graphene by a perpendicular electric field may not be large enough for realistic applications \cite{fior09}. Other methods should be further verified by experiments.

\begin{figure}[!t]
\centering
\includegraphics[width=2.8in]{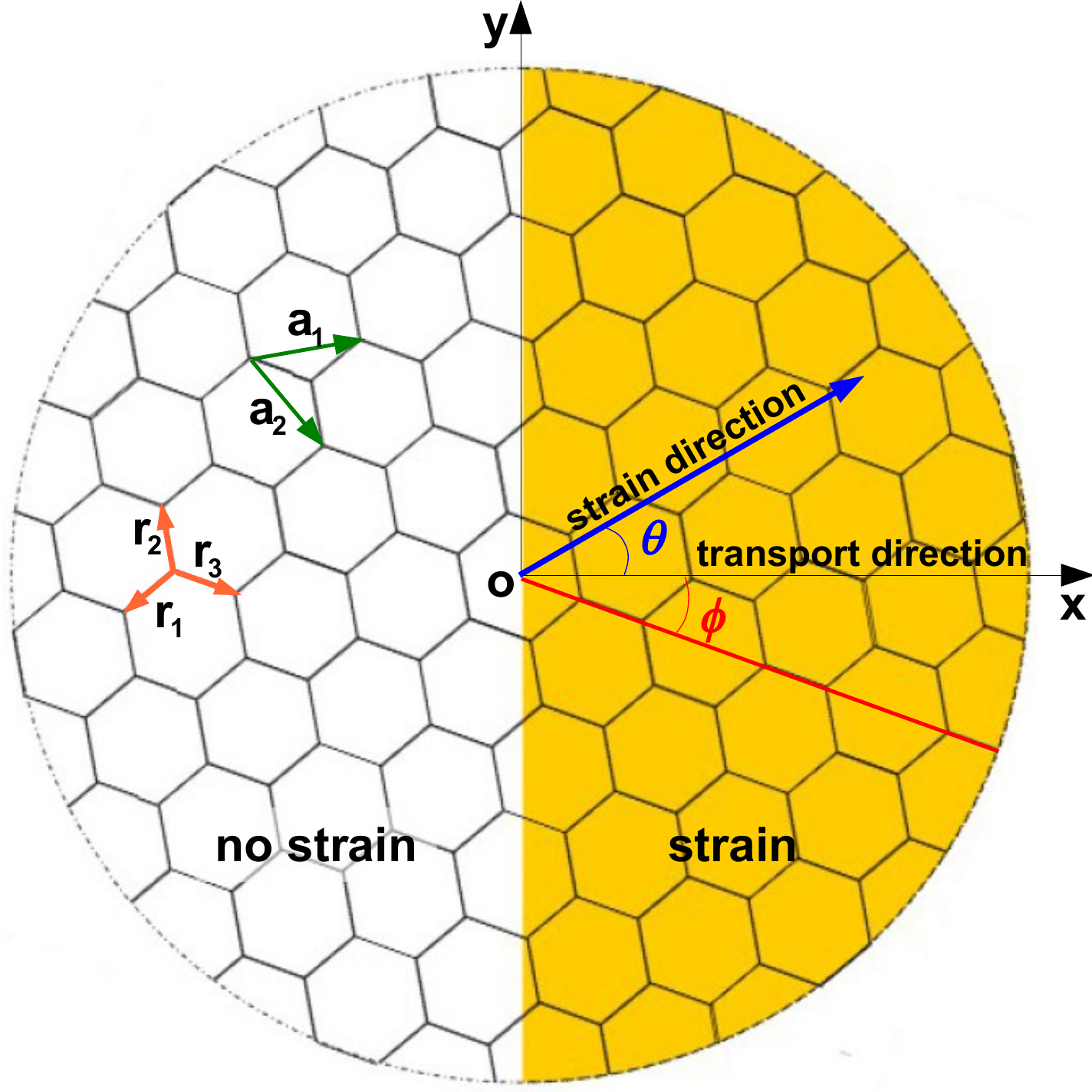}
\caption{Schematic of unstrained/strained graphene junctions investigated in this work.}
\label{fig_sim1}
\end{figure}

On the other hand, graphene was experimentally demonstrated to be able to sustain a much larger strain than conventional semiconductors, making it a promising candidate for flexible electronics (see in a recent review \cite{shar13}). Indeed, strain engineering has been suggested to be an alternative approach to modulating efficiently the electronic properties of graphene nanomaterials. In particular, the bandgap has periodic oscillations in the armchair GNRs \cite{ylu210} while the spin polarization at the ribbon edges (and also the bandgap) can be modulated by the strain in the zigzag cases. In 2D graphene sheets, a finite gap can open under large strains, otherwise, it may remain close to zero but the Dirac points are displaced \cite{cocc10,per209,pere09,huan10}. Many interesting electrical, optical, and magnetic properties induced by strain in graphene have been also explored, e.g. see in \cite{bunc07,pere09,kuma12,per010,pell10,guin10,tlow10,zhai11}.

Besides, local strain is a good option to improve the electrical performance of graphene devices \cite{pere09,ylu010,fuji10,juan11,baha13}. For instance, it has been shown to enhance the ON current in a GNR tunneling FET \cite{ylu010} and to fortify the transport gap in GNR strained junctions \cite{baha13}. In a recent work \cite{hung14}, we have investigated the effects of uniaxial strain on the transport in 2D unstrained/strained graphene junctions and found that due to the strain-induced shift of Dirac points, a significant conduction gap of a few hundreds meV can open with a small strain of a few percent. This type of strained junction was then demonstrated to be an excellent candidate to improve the electronic operation of graphene transistors. It hence motivates us to further investigate the properties of this conduction gap so as to optimize the performance of graphene devices. On the one hand, the effects of strain should be, in principle, dependent on its applied direction. On the other hand, because the appearance of conduction gap is a consequence of the shift of Dirac points along the $k_y$-axis, it is predicted that this gap should also depend on the transport direction. Note that here, Oy (Ox) - axis is assumed to be perpendicular (parallel) to the transport direction. The effects of both strain and transport directions will be clarified systematically in the current work.

\section{Model and calculations}

In this work, the $\pi$-orbital tight binding model constructed in \cite{per209} is used to investigate the electronic transport through the graphene strained junctions schematized in Fig. 1. The Hamiltonian is ${H_{tb}} = \sum\nolimits_{nm} {{t_{nm}}c_n^\dag {c_m}}$ where $t_{nm}$ is the hopping energy between nearest neighbor \emph{n}th and \emph{m}th atoms. The application of a uniaxial strain of angle $\theta$ causes the following changes in the $C-C$ bond vectors:
\begin{eqnarray}
 {{\vec r}_{nm}}\left( \sigma \right) &=& \left\{ {1 + {M_s}\left( \sigma, \theta \right)} \right\}{{\vec r}_{nm}}\left( 0 \right) \\
 {M_s}\left( \sigma, \theta \right) &=& \sigma \left[ {\begin{array}{*{20}{c}}
 {{{\cos }^2}\theta  - \gamma {{\sin }^2}\theta }&{\left( {1 + \gamma } \right)\sin \theta \cos \theta }\\
 {\left( {1 + \gamma } \right)\sin \theta \cos \theta }&{{{\sin }^2}\theta  - \gamma {{\cos }^2}\theta }
 \end{array}} \right] \nonumber
\end{eqnarray}
where $\sigma$ represents the strain and $\gamma \simeq 0.165$ is the Poisson ratio \cite{blak70}. The hopping parameters are defined as $t_{nm} \left( \sigma \right) = t_0 \exp\left[-3.37\left(r_{nm} \left( \sigma \right) /r_0 - 1\right)\right]$, where the hopping energy $t_0 = -2.7$ $eV$ and the bond length $r_{nm} \left( 0 \right) \equiv r_0 = 0.142$ $nm$ in the unstrained case. Therefore, there are three different hoping parameters $t_{1,2,3}$ corresponding to three bond vectors ${\vec r}_{1,2,3}$, respectively, in the strained graphene part of the structure (see Fig. 1). Here, we assume a 1D profile of applied strain, i.e., the strain tensor is a function of position along the transport direction Ox while it is constant along the Oy-axis. The transport direction, $\phi$, and strain direction, $\theta$, are determined as schematized in Fig. 1. Based on this tight binding model, two methods described below can be used to investigate the conduction gap of the considered strained junctions.

\textbf{Green's function calculations.} First, we split the graphene sheet into the smallest possible unit cells periodically repeated along the Ox/Oy directions with the indices $p/q$, respectively (similarly, see the details in \cite{hung12}). The tight-binding Hamiltonian can therefore be expressed in the following form:
\begin{eqnarray}
{H_{tb}} = \sum\limits_{p,q} {\left( {{H_{p,q}} + \sum\limits_{{p_1},{q_1}} {{H_{p,q \to p_1,q_1}}} } \right)}
\end{eqnarray}
where $H_{p,q}$ is the Hamiltonian of cell $\{p,q\}$, and $H_{p,q \to p_1,q_1}$ denotes the coupling of cell $\{p,q\}$ to its nearest neighbor cell $\{p_1,q_1\}$. We then Fourier transform the operators in Eq. (2) as follows:
\begin{eqnarray}
 {c_{p,q}} = \frac{1}{{\sqrt {{M_{cell}}} }}\sum\limits_{{\kappa_y}} {{e^{i{q\kappa_y}}}} {{\hat c}_{p,{\kappa_y}}},
\end{eqnarray}
where $M_{cell}$ is the number of unit cells and $\kappa_y \equiv k_y L_y$ with the size $L_y$ of unit cells along the Oy direction. The Hamiltonian (2) is finally rewritten as a sum of $\kappa_y$-dependent 1D-components:
\begin{eqnarray}
{H_{tb}} &=& \sum\limits_{{\kappa_y}} {\hat H\left( {{\kappa_y}} \right)} \\
\hat H\left( {{\kappa_y}} \right) &=& \sum\limits_p {{{\hat H}_{p \to p - 1}}\left( {{\kappa_y}} \right) + {{\hat H}_p}\left( {{\kappa_y}} \right) + {{\hat H}_{p \to p + 1}}}\left( {{\kappa_y}} \right) \nonumber
\end{eqnarray}
With this Hamiltonian form, the Green's function formalism can be easily applied to compute transport quantities in the graphene strained junction with different transport directions. In particular, the conductance at zero temperature is determined as:
\begin{eqnarray}
 \mathcal{G} \left( \epsilon \right) = \frac{{e^2 W}}{{\pi h L_y}}\int\limits_{BZ} {d{\kappa_y} \mathcal{T}\left( {\epsilon, {\kappa_y}} \right)}
\end{eqnarray}
where $\mathcal{T}\left( {\epsilon,{\kappa_y}} \right)$ is the transmission probability computed from the Green's functions. The integration over $\kappa_y$ is performed in the whole first Brillouin zone. As in ref. \cite{hung13}, the gap of conductance (conduction gap) is then measured from the obtained data of conductance.

\textbf{Bandstructure analyses.} To determine the conduction gap of strained junctions, we find that another simple way based on the analysis of graphene bandstructures could be efficiently used. It is described as follows. Since the conductance is computed from Eq. (5), the appearance of conduction gap is essentially governed by the gaps of transmission probability, which is determined from the energy gaps in the unstrained and strained graphene sections. These energy gaps can be defined directly from the graphene bandstructures. Therefore, our calculation has two steps, similar to that in \cite{hung14}. From the graphene bandstructures obtained using the tight-binding Hamiltonian above, we first look for the energy gaps $E_{unstrain}^{gap}\left( {{\kappa_y}} \right)$ and $E_{strain}^{gap}\left( {{\kappa_y}} \right)$ for a given $\kappa_y$ of two graphene sections. The maximum of these energy gaps determines the gap $E_{junc}^{gap}\left( {{\kappa_y}} \right)$ of transmission probability through the junction. Finally, the conduction gap $E_{cond.gap}$ is obtained by looking for the minimum value of $E_{junc}^{gap}\left( {{\kappa_y}} \right)$ when varying $\kappa_y$ in the whole Brillouin zone.

In particular, the energy bands of strained graphene are given by
\begin{eqnarray}
 E\left( {\vec k} \right) =  \pm \left| {{t_1}{e^{i\vec k{{\vec a}_1}}} + {t_2}{e^{i\vec k{{\vec a}_2}}} + {t_3}} \right|
\end{eqnarray}
where the plus/minus sign corresponds to the conduction/valence bands, respectively. For a given direction $\phi$ of transport, in principle, the vectors $\vec L_{x,y}$ defining the sizes of unit cell along the Ox and Oy directions, respectively, can be always expressed as ${\vec L_x} = {n_1}{\vec a_1} + {n_2}{\vec a_2}$ and ${\vec L_y} = {m_1}{\vec a_1} + {m_2}{\vec a_2}$ with $\cos \phi = \frac{{{{\vec L}_x}\vec L_x^0}}{{{L_x}L_x^0}}$ and $\sin \phi = \frac{{{{\vec L}_x}\vec L_y^0}}{{{L_x}L_y^0}}$ while $\vec L_{x,y}^0 = {\vec a_1} \pm {\vec a_2}$. Note that $n_{1,2}$ and $m_{1,2}$ are integers while $\frac{{{m_1}}}{{{m_2}}} =  - \frac{{{n_1} + 2{n_2}}}{{{n_2} + 2{n_1}}}$, i.e., ${\vec L_{x}} {\vec L_{y}} = 0$. In other words, we have the following expressions
\begin{eqnarray}
{{{\vec a}_1} = \frac{{ - {m_2}{{\vec L}_x} + {n_2}{{\vec L}_y}}}{{{n_2}{m_1} - {n_1}{m_2}}};\,\,\,{{\vec a}_2} = \frac{{{m_1}{{\vec L}_x} - {n_1}{{\vec L}_y}}}{{{n_2}{m_1} - {n_1}{m_2}}}}
\end{eqnarray}
On this basis, the energy bands can be rewritten in terms of $\kappa_{x, y} = \vec k \vec L_{x,y} \left( { \equiv {k_{x,y}}{L_{x,y}}} \right)$ by substituting Eqs. (7) into Eq. (6). This new form of energy bands is finally used to compute the conduction gap of strained junctions.

As a simple example, in the case of $\phi = 0$ (armchair direction), we calculate the conduction gap as follows. First, Eq. (6) is rewritten in the form
\begin{eqnarray}
 E_{\phi = 0}\left( {\vec \kappa} \right) =  \pm \left| {{t_1}{e^{i\kappa_y/2}} + {t_2}{e^{ - i\kappa_y/2}} + {t_3}{e^{ - i\kappa_x/2}}} \right|
\end{eqnarray}
with the vectors $\vec L_{x,y} \equiv \vec L_{x,y}^0$. Using this new form, the energy gap of strained graphene for a given $\kappa_y$ is determined as
\begin{equation}
{E_{strain}^{gap}}\left( {{\kappa_y}} \right) = 2 \left| {\sqrt {{{\left( {{t_1} - {t_2}} \right)}^2} + 4{t_1}{t_2}{{\cos }^2}\frac{{{\kappa_y}}}{2}}  + {t_3}} \right|
\end{equation}
while ${E_{unstrain}^{gap}}\left( {{\kappa_y}} \right)$ is given by the same formula with $t_1$ = $t_2$ = $t_3$ $\equiv$ $t_0$. The gap of transmission probability through the junction is then determined as ${E_{junc}^{gap}}\left( {{\kappa_y}} \right) = \max \left[ {E_{unstrain}^{gap}\left( {{\kappa_y}} \right),E_{strain}^{gap}\left( {{\kappa_y}} \right)} \right]$ and, finally, the conduction gap is given by ${E_{cond.gap}} = \min \left[ {E_{junc}^{gap}\left( {{\kappa_y}} \right)} \right]$ for $\kappa_y$ in the whole Brillouin zone.

We would like to notice that the Green's function calculations and the banstructure analyses give the same results of conduction gap in the junctions where the transition region between unstrained and strained graphene sections is long enough, i.e., larger than about 5 to 6 nm. In the case of short length, as discussed in \cite{baha13,hung14}, this transition zone can have significant effects on the transmission between propagating states beyond the energy gaps and hence can slightly enlarge the gap of conductance, compared to the results obtained from the bandstructure calculations.

\section{Results and discussion}

\begin{figure}[!t]
\centering
\includegraphics[width=3.0in]{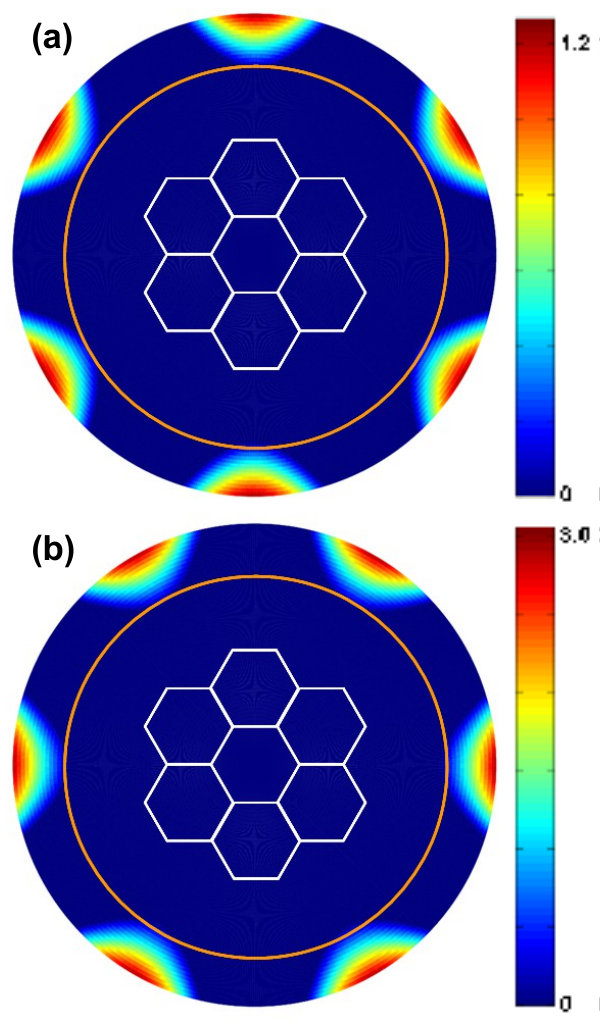}
\caption{Dependence of graphene bandgap (in the unit of eV) on the applied strain and its direction: tensile (a) and compressive (b). The radius from the central point indicates the strain strength ranging from 0 (center) to 30 $\%$ (edge of maps) while the graphene lattice is superimposed to show visibly the strain direction. The orange circle corresponds to the strains of $\sigma = 23 \%$.}
\label{fig_sim2}
\end{figure}
\begin{figure}[!t]
\centering
\includegraphics[width=3.4in]{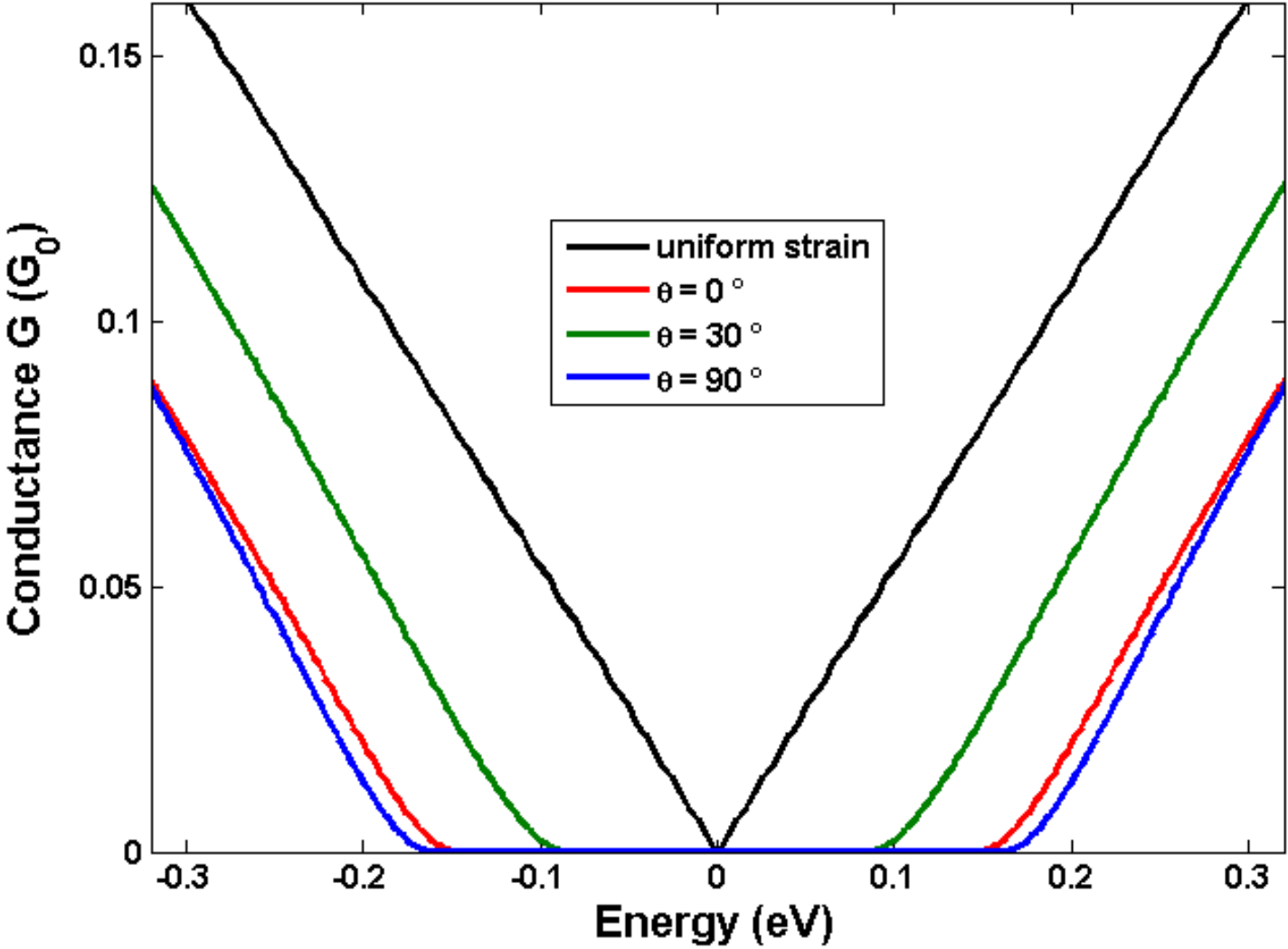}
\caption{Conductance ($G_0 = e^2W/hL_y$) as a function of energy in graphene strained junctions for $\sigma = 4 \%$ with different strain directions. The transport along the armchair direction ($\phi = 0$) is considered. The data obtained in a uniformly strained graphene is displayed for the comparison.}
\label{fig_sim6}
\end{figure}
\begin{figure*}[!t]
\centering
\includegraphics[width=5.8in]{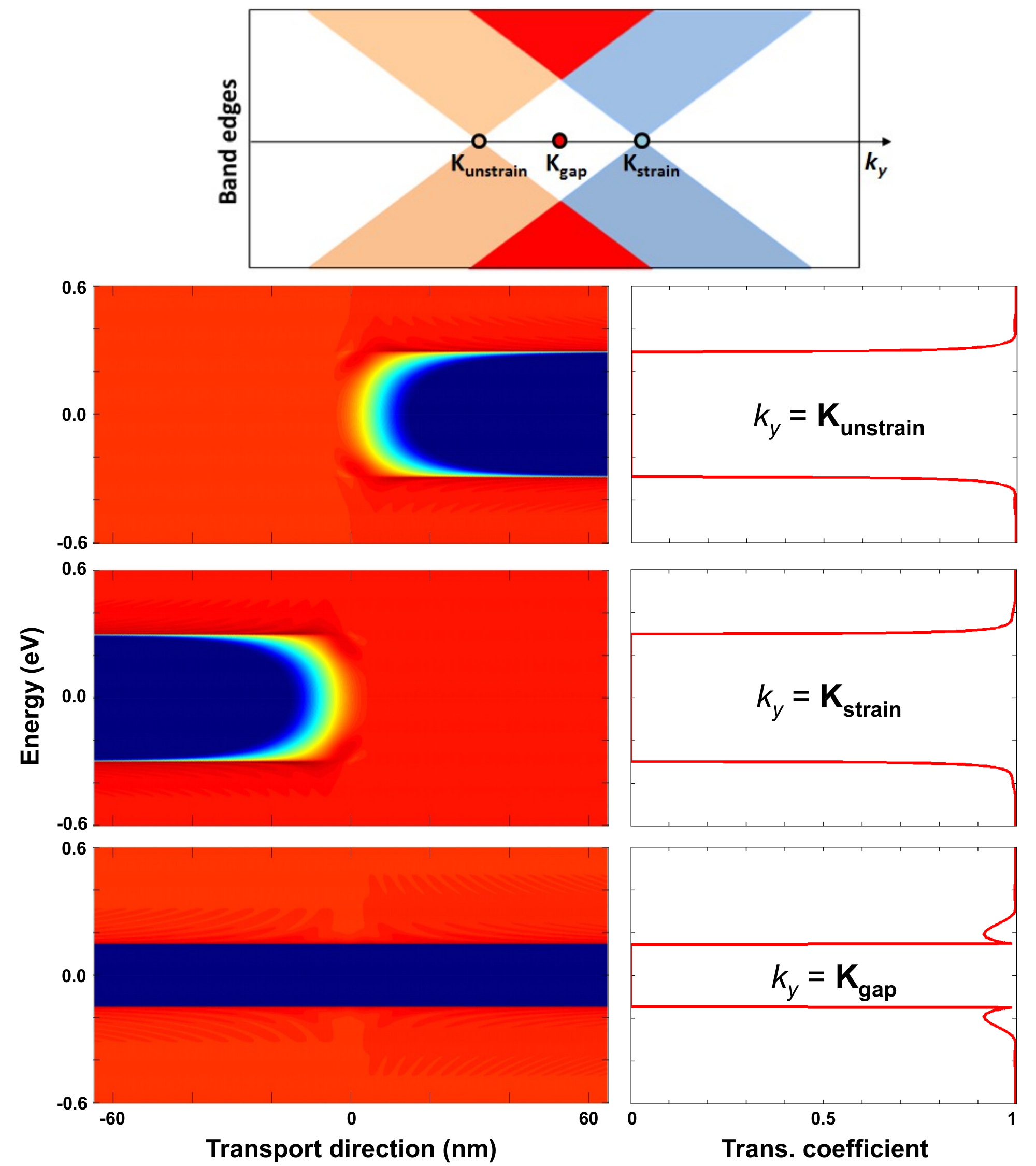}
\caption{Local density of states (left panels) and corresponding transmission coefficient (right panels) for three different wave-vectors $k_y$ obtained in an unstrained/strained graphene junction of $\sigma = 4 \%$, and $\theta \equiv \phi = 0$. On the top is a schematic of graphene bandedges illustrating the strain-induced shift of Dirac points along the $k_y$-direction.}
\label{fig_sim4}
\end{figure*}
\begin{figure*}[!t]
\centering
\includegraphics[width=5.6in]{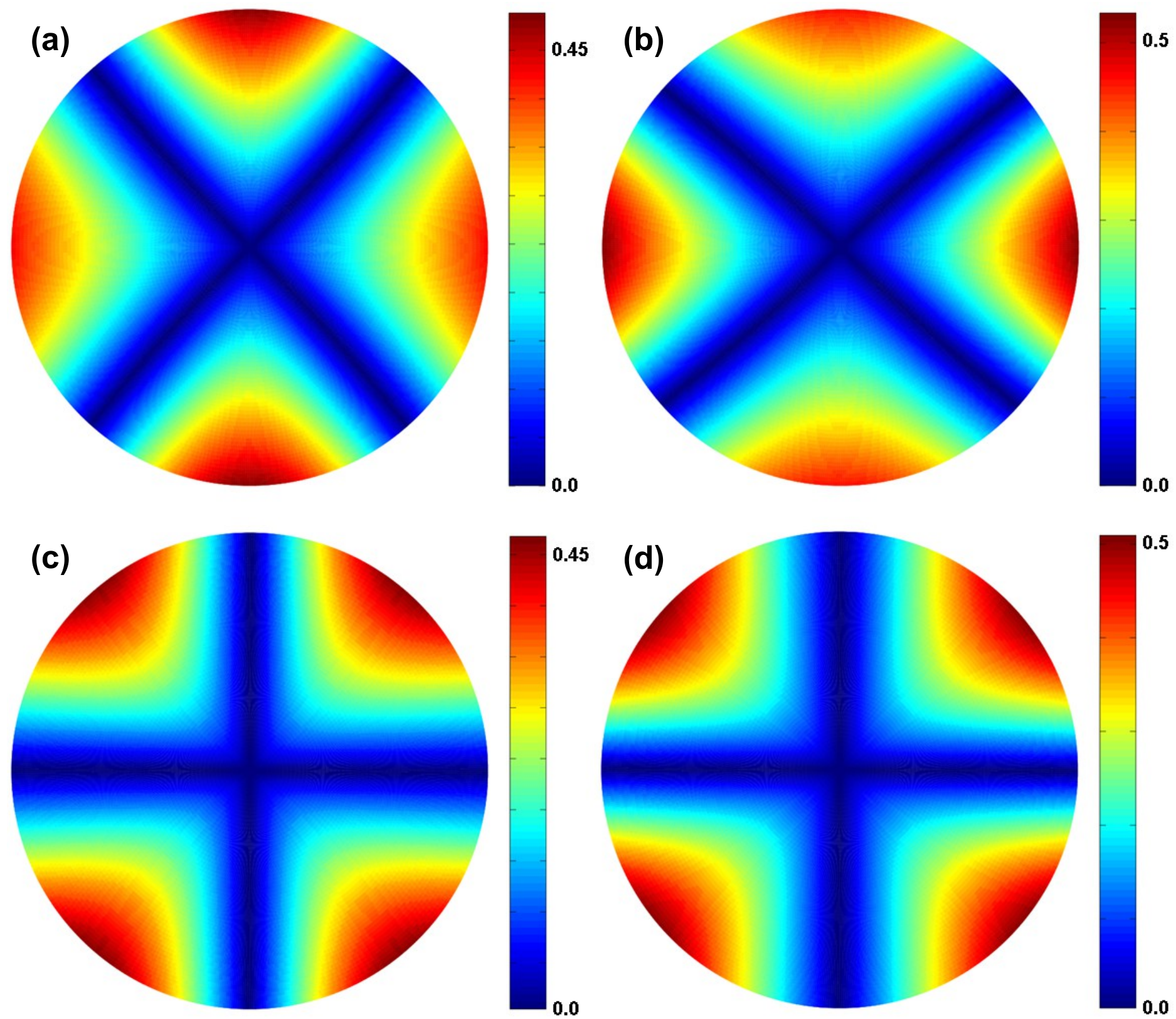}
\caption{Maps of conduction gap in unstrained/strained graphene junctions: tensile (a,c) and compressive cases (b,d). The transport is along the armchair $\phi = 0$ (a,b) and zigzag $\phi = 30^\circ$ directions (c,d). The strain strength ranges from 0 (center) to 6 $\%$ (edge of maps) in all cases.}
\label{fig_sim4}
\end{figure*}
% \begin{figure}[!t]
% \centering
% \includegraphics[width=3.3in]{Fig06.pdf}
% \caption{Conduction gap as a function of strain direction with $\sigma = 4 \%$. Different transport directions are investigated.}
% \label{fig_sim6}
% \end{figure}
First, we re-examine the formation of the bandgap of graphene under a uniaxial strain. From Eq. (9), it is shown that a strain-induced finite-bandgap appears only if ${E_{strain}^{gap}}\left( {{\kappa_y}} \right) > 0$ for all $k_y$ in the first Brillouin zone, i.e., ${k _y} \in \left[ { - \frac{\pi}{L_y}, \frac{\pi}{L_y}} \right]$, otherwise, the bandgap remains zero. Hence, the condition for the bandgap to be finite is either
\begin{equation*}
 \left| {{t_1} - {t_2}} \right| > \left| {{t_3}} \right|\,\,\,\,\,{\rm{OR}}\,\,\,\,\,\left| {{t_3}} \right| > \left| {{t_1} + {t_2}} \right|
\end{equation*}
and the corresponding values of bandgap are
\begin{equation*}
 {E_{gap}} = 2\left( {\left| {{t_1} - {t_2}} \right| - \left| {{t_3}} \right|} \right)\,\,\,\,\,{\rm{OR}}\,\,\,\,\,2\left( {\left| {{t_3}} \right| - \left| {{t_1} + {t_2}} \right|} \right)
\end{equation*}
This result was actually reported in \cite{per209,hase06}. We remind as displayed in Fig. 2(a) that a finite bandgap opens only for strain larger than $\sim 23 \%$ and the zigzag (not armchair) is the preferred direction for bandgap opening under a tensile strain \cite{per209}. We extend our investigation to the case of compressive strain and find  (see in Fig. 2(b)) that (i) the same gap threshold of $\sigma \simeq 23 \%$ is observed but (ii) the preferred direction to open the gap under a compressive strain is the armchair, not the zigzag as the case of tensile strain. This implies that the properties of graphene bandstructure at low energy should be qualitatively the same when applying strains of $\left\{ {\sigma ,\theta } \right\}$ and of $\left\{ {-\sigma ,\theta + 90^\circ} \right\}$. This feature can be understood by considering, for example, strains of $\left\{ {\sigma , \theta = 0} \right\}$ and of $\left\{ {-\sigma , \theta = 90^\circ} \right\}$. Indeed, these strains result in the same qualitative changes on the bond-lengths, i.e., an increased bond-length $r_3$ and reduced bond-lengths $r_{1,2}$. However, for the same strain strength, because of the exponential dependence of hoping energies on the bond-lengths, the compressive strain generally induces a larger bandgap than the tensile one, as can be seen when comparing the data displayed in Figs. 2(a) and 2(b). To conclude, we would like to emphasize that a large strain is necessary to open a bandgap in graphene. This could be an issue for practical applications, compared to the use of graphene strained junctions explored in \cite{hung14}.

We now go to explore the properties of conduction gap in the graphene strained junctions. In Fig. 3, we display the conductance as a function of energy computed from Eq. (5) using the Green's function technique. As discussed above, a small strain of a few percent (e.g., 4 $\%$ here) can not change the gapless character of graphene, i.e., there is no gap of conductance in the case of uniformly strained graphene. However, similar to that reported in \cite{hung14}, a significant conduction-gap of a few hundreds meV can open in the unstrained/strained graphene junctions. The appearance of this conduction gap, as mentioned previously, is due to the strain-induced shift of Dirac points and is explained as follows. Actually, the strain causes the lattice deformation and can result in the deformation of graphene bandstructure. Therefore, the bandedges as a function of wave-vector $k_y$ in unstrained and strained graphene can be illustrated schematically as in the top panel of Fig. 4. As one can see, the shift of Dirac points leads to the situation where there is no value of $\kappa_y$, for which the energy gaps $E_{unstrain}^{gap}\left( {{\kappa_y}} \right)$ and $E_{strain}^{gap}\left( {{\kappa_y}} \right)$ are simultaneously equal to zero. This means that the transmission probability always shows a finite gap for any $\kappa_y$. For instance, the energy gap is zero (or small) in the unstrained (resp. strained) graphene section but finite in the strained (resp. unstrained) one in the vicinity of Dirac point $k_y = K_{unstrain}$ (resp. $K_{strain}$). Accordingly, as illustrated in the pictures of LDOS in the left panels of Fig. 4 and confirmed in the corresponding transmissions in the right panels, clear gaps of transmission are still obtained. Far from these values of $k_y$, $E_{unstrain}^{gap}\left( {{\kappa_y}} \right)$ and $E_{strain}^{gap}\left( {{\kappa_y}} \right)$ are both finite (e.g., see the LDOS plotted for $k_y = K_{gap}$) and hence a finite gap of transmission also occurs. On this basis, a finite gap of conductance is achieved. More important, Fig. 3 shows that besides the strength of strain, the strain effect is also strongly dependent on the applied direction. For instance, the conduction gap takes the values of $\sim$ 295, 172 and 323 meV for $\theta = 0$, $30^\circ$ and $90^\circ$, respectively.

Below, we will discuss the properties of the conduction gap with respect to the strain, its applied direction, and the direction of transport. Note that due to the lattice symmetry, the transport directions $\phi$ and $\phi + 60^\circ$ are equivalent while the applied strain of angle $\theta$ is identical to that of $\theta + 180^\circ$. Hence, the data obtained for $\phi$ ranging from $-30^\circ$ to $30^\circ$ and $\theta  \in \left[ {0^\circ ,180^\circ } \right]$ covers the properties of conduction gap in all possible cases.

In Fig. 5, we present the maps of conduction gap with respect to the strain and its applied direction in two particular cases: the transport is either along the armchair ($\phi = 0$) or the zigzag ($\phi = 30^\circ$) directions. Both tensile and compressive strains are considered. Let us first discuss the results obtained in the armchair case. Figs. 5(a,b) show that (i) a large conduction gap up to about 500 meV can open with a strain of 6 $\%$ and (ii) again the conduction gap is strongly $\theta$-dependent, in particular, its peaks occur at $\theta = 0$ or $90^\circ$ while the gap is zero at $\theta \approx 47^\circ$ and $133^\circ$ for tensile strain and at $\theta \approx 43^\circ$ and $137^\circ$ for compressive strain. In principle, the conduction gap is larger if the shift of Dirac points in the $\kappa_y$-axis is larger, as discussed above about Figs. 3-4. We notice that the strain-induced shifts can be different for the six Dirac points of graphene \cite{kitt12} and the gap is zero when there is any Dirac point observed at the same $\kappa_y$ in the two graphene sections. From Eq. (9), we find that the Dirac points are determined by the following equations:
\begin{eqnarray*}
  {\cos}\frac{\kappa_y}{2} &=& \pm \frac{1}{2}\sqrt{\frac{{t_3^2 - {{\left( {{t_1} - {t_2}} \right)}^2}}}{{{t_1}{t_2}}}}, \\
  \cos \frac{{\kappa_x}}{2} &=& \frac{{{t_1} + {t_2}}}{{\left| {{t_3}} \right|}}\cos \frac{{\kappa_y}}{2},\,\,\,\sin \frac{{\kappa_x}}{2} = \frac{{{t_2} - {t_1}}}{{\left| {{t_3}} \right|}}\sin \frac{{\kappa_y}}{2},
\end{eqnarray*}
which simplify into ${\cos}\frac{\kappa_y}{2} = \pm \frac{1}{2}$ and, respectively, $\cos \left( {\frac{{{\kappa _x}}}{2}} \right) = \mp 1$ in the unstrained case. Hence, the zero conduction gap is obtained if
\begin{equation*}
  \frac{{t_3^2 - {{\left( {{t_1} - {t_2}} \right)}^2}}}{{4{t_1}{t_2}}} = \frac{1}{4}
\end{equation*}
Additionally, it is observed that the effects of a strain $\{\sigma,\theta\}$ are qualitatively similar to those of a strain $\{-\sigma,\theta+90^\circ\}$, i.e., the peaks and zero values of conduction gap are obtained at the same $\theta$ in these two situations. To understand this, we analyze the strain matrix $M_s \left(\sigma,\theta\right)$ and find that in the case of small strains studied here, there is an approximate relationship between the bond lengths under these two strains, given by \[{r \left( \sigma, \theta \right)} - {r \left( -\sigma, \theta + 90^\circ\right)} \simeq \sigma \left( {1 - \gamma } \right) r_0,\] which is $\theta$-independent for all \emph{C-C} bond vectors. It implies that there is a fixed ratio between the hopping energies $t_i \left( \sigma, \theta \right)$ and $t_i \left( -\sigma, \theta + 90^\circ\right)$ and hence there is the similar shift of Dirac points in these two cases.
\begin{figure}[!t]
\centering
\includegraphics[width=3.4in]{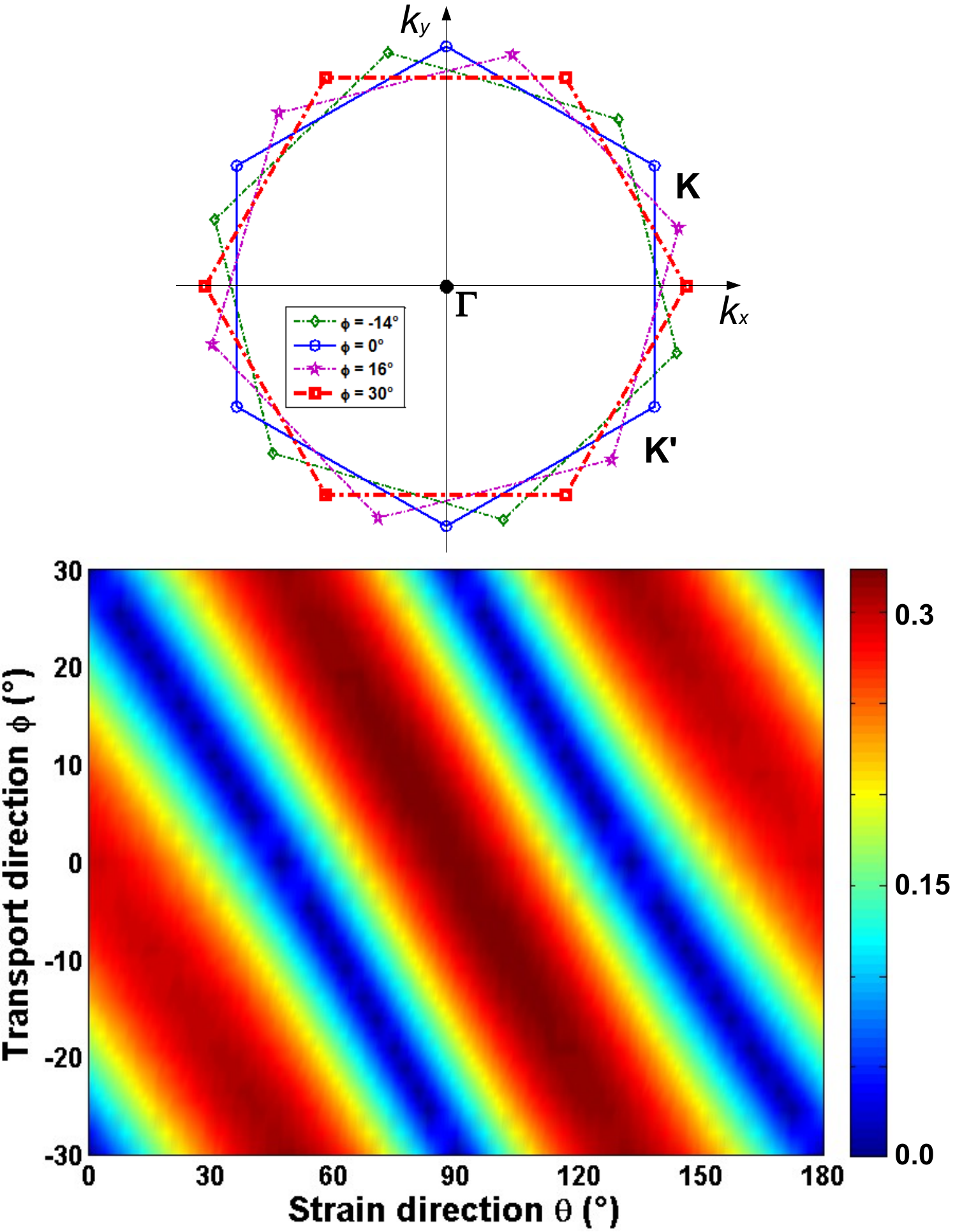}
\caption{Map showing the dependence of conduction gap on the directions ($\theta,\phi$) for $\sigma = 4 \%$. The top is a diagram illustrating the rotation of Dirac points in the \emph{k}-space with the change in the transport direction $\phi$.}
\label{fig_sim6}
\end{figure}
\begin{figure*}[!t]
\centering
\includegraphics[width=5.5in]{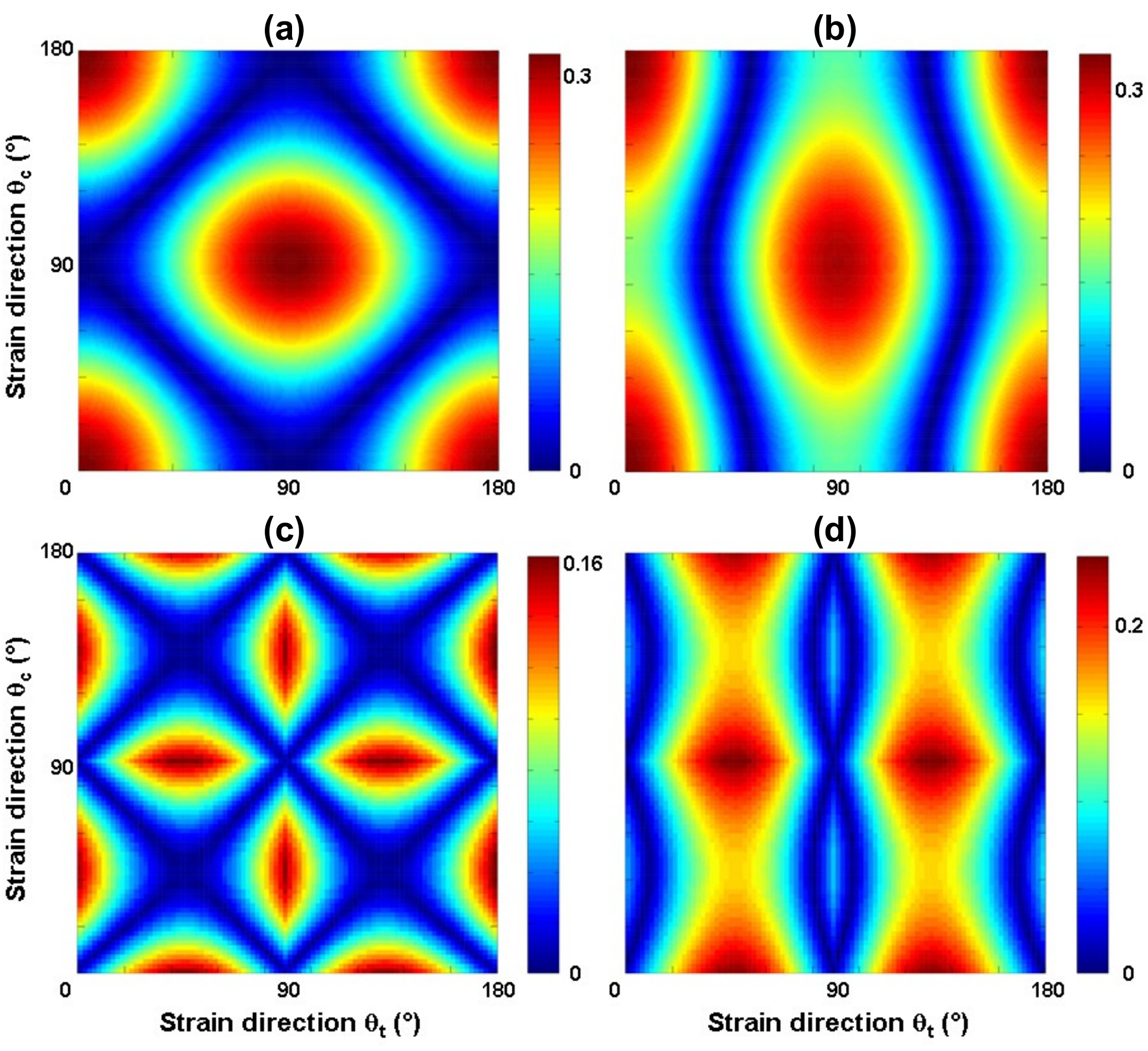}
\caption{Maps of conduction gap obtained in tensile/compressive strained junctions. The transport along the armchair/zigzag directions is considered in (a,b)/(c,d), respectively. The strains $\sigma_c = -2 \%$ and $\sigma_t = 2 \%$ are applied in (a,c) while $\sigma_c = -1 \%$ and $\sigma_t = 3 \%$ in (b,d).}
\label{fig_sim4}
\end{figure*}

We now go to analyze the properties of conduction gap shown in Figs. 5(c,d) where the transport is along the zigzag direction $\phi = 30^\circ$. In fact, the conduction gap in this case can reach a value as high as that of the case of $\phi = 0$ but has different $\theta$-dependence. In particular, the conduction gap has peaks at $\theta \approx 47^\circ$ and $133^\circ$ for tensile strain and at $\theta \approx 43^\circ$ and $137^\circ$ for compressive strain, where it is zero in the case of $\phi = 0$. It is also equal to zero at $\theta = 0$ and $\theta = 90^\circ$ where the peaks of conduction gap occur in the latter case of $\phi = 0$. The relationship between these two transport directions can be explained as follows. On the one hand, based on the analyses above for $\phi = 0$, we find that for a given strength of strain, a maximum shift of Dirac points along the $k_y$-axis corresponds to a minimum along the $k_x$-one and vice versa when varying the strain direction $\theta$. On the other hand, as schematized in the top of Fig. 6 below, the change in the transport direction results in the rotation of the first Brillouin zone, i.e., the $k_x$ (resp. $k_y$) axis in the case of $\phi = 30^\circ$ is identical to the $k_y$ (resp. $k_x$) axis in the case of $\phi = 0$. These two features explain essentially the opposite $\theta$-dependence of conduction gap for $\phi = 30^\circ$, compared to the case of $\phi = 0$ as mentioned. Again, we found the same qualitative behavior of conduction gap when applying the strains of $\{\sigma,\theta\}$ and $\{-\sigma,\theta+90^\circ\}$.

Next, we investigate the conduction gap with respect to different transport directions $\phi$. We display a ($\theta,\phi$)-map of conduction gap for $\sigma = 4 \%$ in Fig. 6 and, in the top, an additional diagram illustrating the rotation of Dirac points in the $k-$space with the change in the transport direction. It is clearly shown that (i) a similar scale of conduction gap is obtained for all different transport directions, (ii) there is a smooth and continuous shift of $E_{cond.gap}-\theta$ behavior when varying $\phi$, and (iii) the same behavior of $E_{cond.gap}$ is also observed when comparing the two transport directions of $\phi$ and $\phi+30^\circ$, similarly to the comparison above between $\phi = 0^\circ$ and $30^\circ$. The data plotted in Fig. 6 additionally shows that $E_{cond.gap}$ takes the same value in both cases of $\{\phi,\theta\}$ and $\{-\phi,-\theta\}$ with a remark that the strains of $-\theta$ and $180^\circ-\theta$ are identical. Moreover, the values of $\theta$ and $\phi$, for which the conduction gap has a peak or is equal to zero, have an almost linear relationship. In particular, the relationship for conduction gap peaks is approximately given by $\theta = \theta_A - \eta_s \phi$. For tensile strains, $\eta_s$ takes the values of $\sim 1.5667$ and $1.4333$ for $\theta_A = 0$ and $90^\circ$, respectively. On the opposite, it is about $1.4333$ and $1.5667$ for $\theta_A = 0$ and $90^\circ$, respectively, for compressive strain cases. All these features are consequences of the rotation of Dirac points in the $k$-space with respect to the transport direction $\phi$ as illustrated in the diagram on the top and the lattice symmetry of graphene.

Finally, we investigate other junctions based on compressive and tensile strained graphene sections. The idea is that in this type of strained junction, the shifts of Dirac points are different in two graphene sections of different strains, which offers the possibilities to use smaller strains to achieve a similar conduction gap, compared to the case of unstrained/strained junction. In Fig. 7, we display the maps of conduction gap with respect to the directions of compressive ($\theta_c$) and tensile ($\theta_t$) strains in two cases of transport direction $\phi = 0$ (armchair) and $30^\circ$ (zigzag) for given strain strengths. Indeed, as seen in Fig. 7(a,b), with smaller strains $\left\{ {{\sigma _c},{\sigma _t}} \right\} = \left\{ { - 2\% ,2\% } \right\}$ or $\left\{ { - 1\% ,3\% } \right\}$, similar conduction gap of about 310 meV can be achieved (see Figs. 7(a,b)) while it requires a strain of 4 $\%$ in the unstrained/strained junctions discussed above. However, since the shift of Dirac points is strongly dependent on the direction of applied strains and the transport direction, the properties of conduction gap are more complicated than in the latter case. In particular, our calculations show that the preferred transport directions to achieve a large conduction gap are close to the armchair one. Otherwise, the conduction gap is generally smaller, similarly to the data for $\phi = 30^\circ$ compared to $\phi = 0$, as shown in Fig. 7. Additionally, it is shown that the preferred directions of applied strains in the case of $\phi = 0$ are close to ${\theta _c} \equiv {\theta _t} = 0$ or $90^\circ$.

\section{Conclusion}

Based on the tight binding calculations, we have investigated the effects of uniaxial strain on the transport properties of graphene strained junctions and discuss systematically the possibilities of achieving a large conduction gap with respect to the strain, its applied direction and the transport direction. It has been shown that due to the strain-induced deformation of graphene lattice and hence of graphene bandstructure, a finite conduction gap higher than 500 meV can be achieved for a strain of only 6 $\%$. Moreover, as a consequence of the shift of Dirac points along the $k_y$-axis, the conduction gap is strongly dependent not only on the strain strength but also on the direction of applied strain and the transport direction. A full picture of these properties of conduction gap has been presented and explained. The study hence could be a good guide for the use of this type of unstrained/strained graphene junction in electronic applications.

\textbf{\textit{Acknowledgment.}} This research in Hanoi is funded by Vietnam National Foundation for Science and Technology Development (NAFOSTED) under grant number 103.02-1012.42. We also acknowledges the French ANR for financial support under the projects NANOSIM-GRAPHENE (Grant no. ANR-09-NANO-016) and MIGRAQUEL (Grant no. ANR-10-BLAN-0304).


\begin{thebibliography}{99}
\bibitem{ahcn09} A. H. Castro Neto, F. Guinea, N. M. R. Peres, K. S. Novoselov, and A. K. Geim, Rev. Mod. Phys. \textbf{81}, 109 (2009).
\bibitem{schw10} F. Schwierz, Nat. Nanotechnol. \textbf{5}, 487 (2010).
\bibitem{ywu013} Y. Wu, D. B. Farmer, F. Xia, and P. Avouris, Proc. IEEE \textbf{101}, 1620 (2013).
\bibitem{bolo08} K. I. Bolotin, K. J. Sikes, J. Jiang, M. Klima, G. Fudenberg, J. Hone, P. Kim, and H. L. Stormer, Solid State Commun. \textbf{146}, 351 (2008).
\bibitem{zome11} P. J. Zomer, S. P. Dash, N. Tombros, and B. J. van Wees, Appl. Phys. Lett. \textbf{99}, 232104 (2011).
\bibitem{novo04} K. S. Novoselov, A. K. Geim, S. V. Morozov, D. Jiang, Y. Zhang, S. V. Dubonos, I. V. Grigorieva, and A. A. Firsov, Science \textbf{306}, 666 (2004).
\bibitem{ywu012} Y. Wu, K. A. Jenkins, A. Valdes-Garcia, D. B. Farmer, Y. Zhu, A. A. Bol, C. Dimitrakopoulos, W. Zhu, F. Xia, Ph. Avouris, and Y.-M. Lin, Nano Lett. \textbf{12}, 3062 (2012).
\bibitem{chen12} R. Cheng, J. Bai, L. Liao, H. Zhou, Y. Chen, L. Liu, Y. C. Lin, S. Jiang, Y. Huang, and X. Duan, Proc. Nat. Acad. Sci. USA \textbf{109}, 11588 (2012).
\bibitem{zguo13} Z. Guo, R. Dong, P. S. Chakraborty, N. Lourenco, J. Palmer, Y. Hu, M. Ruan, J. Hankinson, J. Kunc, J. D. Cressler, C. Berger, and W. A. de Heer, Nano Lett. \textbf{13}, 942 (2013).
\bibitem{meri08} I. Meric, M. Y. Han, A. F. Young, B. Ozyilmaz, P. Kim, and K. L. Shepard, Nat. Nanotechnol. \textbf{3}, 654, (2008).
\bibitem{yhan07} M. Y. Han, B. Ozyilmaz, Y. Zhang, and P. Kim, Phys. Rev. Lett. \textbf{98}, 206805 (2007).
\bibitem{khar11} N. Kharche and S. K. Nayak, Nano Lett. \textbf{11}, 5274 (2011).
\bibitem{lher13} A. Lherbier, A. R. Botello-Mendez, and J.-C. Charlier, Nano Lett. \textbf{13}, 1446 (2013).
\bibitem{jbai10} J. Bai, X. Zhong, S. Jiang, Y. Huang, and X. Duan, Nat. Nanotechnol. \textbf{5}, 190 (2010).
\bibitem{berr13} S. Berrada, V. Hung Nguyen, D. Querlioz, J. Saint-Martin, A. Alarc\'{o}n, C. Chassat, A. Bournel, and P. Dollfus, Appl. Phys. Lett. \textbf{103}, 183509 (2013).
\bibitem{zhan09} Y. Zhang, T. T. Tang, C. Girit, Z. Hao, M. C. Martin, A. Zettl, M. F. Crommie, Y. Ron Shen, and F. Wang, Nature 459, 820 (2009).
\bibitem{quer08} D. Querlioz, Y. Apertet, A. Valentin, K. Huet, A. Bournel, S. Galdin-Retailleau, and P. Dollfus, Appl. Phys. Lett. \textbf{92}, 042108 (2008).
\bibitem{hung13} V. Hung Nguyen, M. Chung Nguyen, H. Viet Nguyen, and P. Dollfus, J. Appl. Phys. \textbf{113}, 013702 (2013).
\bibitem{fior09} G. Fiori and G. Iannaccone, IEEE Electron Device Lett. \textbf{30}, 261 (2009).

\bibitem{shar13} B. K. Sharma and J.-H. Ahn, Solid State Electron. \textbf{89}, 177 (2013).
%\bibitem{clee08} C. Lee, X. Wei, J. W. Kysar, and J. Hone, Science \textbf{321}, 385 (2008).
\bibitem{ylu210} Y. Lu and J. Guo, Nano Res. \textbf{3}, 189 (2010).
\bibitem{cocc10} G. Cocco, E. Cadelano, and L. Colombo, Phys. Rev. B \textbf{81}, 241412 (2010).
\bibitem{per209} V. M. Pereira and A. H. Castro Neto, and N. M. R. Peres, Phys. Rev. B \textbf{80}, 045401 (2009).
\bibitem{huan10} M. Huang, H. Yan, T. F. Heinz, and J. Hone, Nano Lett. \textbf{10}, 4074 (2010).
\bibitem{pere09} V. M. Pereira and A. H. Castro Neto, Phys. Rev. Lett. \textbf{103}, 046801 (2009).

\bibitem{bunc07} J. S. Bunch, A. M. van der Zande, S. S. Verbridge, I. W. Frank, D. M. Tanenbaum, J. M. Parpia, H. G. Craighead, P. L. McEuen, Science \textbf{315}, 490 (2007).
\bibitem{kuma12} S. Bala Kumar and J. Guo, Nano Lett. \textbf{12}, 1362 (2012).
\bibitem{per010} V. M. Pereira, A. H. Castro Neto, H. Y. Liang and L. Mahadevan, Phys. Rev. Lett. \textbf{105}, 156603 (2010).
\bibitem{pell10} F. M. D. Pellegrino, G. G. N. Angilella, and R. Pucci, Phys. Rev. B \textbf{81}, 035411 (2010).
\bibitem{guin10} F. Guinea, M. I. Katsnelson, and A. K. Geim, Nat. Phys. \textbf{6}, 30 (2010).
\bibitem{tlow10} T. Low and F. Guinea, Nano Lett. \textbf{10}, 3551 (2010).
\bibitem{zhai11} F. Zhai and L. Yang, Appl. Phys. Lett. \textbf{98}, 062101 (2011)

\bibitem{ylu010} Y. Lu and J. Guo, App. Phys. Lett. \textbf{97}, 073105 (2010).
\bibitem{fuji10} T. Fujita, M. B. A. Jalil, and S. G. Tan, Appl. Phys. Lett. \textbf{97}, 043508 (2010).
\bibitem{juan11} F. de Juan, A. Cortijo, M. A. H. Vozmediano and A. Cano, Nat. Phys. \textbf{7}, 810 (2011).
\bibitem{baha13} D. A. Bahamon and V. M. Pereira, Phys. Rev. B \textbf{88}, 195416 (2013).

\bibitem{hung14} V. Hung Nguyen, H. Viet Nguyen, and P. Dollfus, Nanotechnol. \textbf{25}, 165201 (2014).

\bibitem{blak70} O. L. Blakslee, D. G. Proctor, E. J. Seldin, G. B. Spence and T. Weng, J. Appl. Phys. \textbf{41}, 3373 (1970).
\bibitem{hung12} V. Hung Nguyen, F. Mazzamuto, J. Saint-Martin, A. Bournel, and P. Dollfus, Nanotechnol. \textbf{23}, 065201 (2012).

\bibitem{hase06} Y. Hasegawa, R. Konno, H. Nakano, and M. Kohmoto, Phys. Rev. B \textbf{74}, 033413 (2006).
\bibitem{kitt12} A. L. Kitt, V. M. Pereira, A. K. Swan, and B. B. Goldberg, Phys. Rev. B \textbf{85}, 115432 (2012).

\end{thebibliography}
\end{document}